\documentclass[12pt]{amsart}

\usepackage{graphicx}
\usepackage{amsmath}

\newtheorem{thm}{Theorem}
\newtheorem{lemma}[thm]{Lemma}
\newtheorem{prop}[thm]{Proposition}
\newtheorem{cor}[thm]{Corollary}
\theoremstyle{definition}
\newtheorem{defn}[thm]{Definition}
\newtheorem{example}[thm]{Example}
\newtheorem{remark}[thm]{Remark}
\newtheorem{conj}[thm]{Conjecture}

\newcommand{\ignore}[1]{}

\begin{document}

\title{Why neighbor-joining works}
\author{Radu Mihaescu \and Dan Levy \and Lior Pachter}
\address{Department of Mathematics and Computer Science, UC Berkeley}
\email{\{mihaescu,levyd,lpachter\}@math.berkeley.edu}

\begin{abstract}
We show that the neighbor-joining algorithm is a robust quartet
method for constructing trees from distances. This leads to a new
performance guarantee that contains Atteson's optimal radius bound
as a special case and explains many cases where neighbor-joining
is successful even when Atteson's criterion is not satisfied. We
also provide a proof for Atteson's conjecture on the optimal edge
radius of the neighbor-joining algorithm. The strong performance
guarantees we provide also hold for the quadratic time fast
neighbor-joining algorithm, thus providing a theoretical basis for
inferring very large phylogenies with neighbor-joining.
\end{abstract}


\maketitle

\section{Introduction}

The widely used neighbor-joining algorithm \cite{Saitou1987} has
been extensively analyzed and compared to other tree construction
methods. Previous studies have mostly focused on empirical testing
of neighbor-joining. Examples include the comparison of
neighbor-joining with quartet \cite{StJohn2003} and maximum
likelihood \cite{Huelsenbeck1993} methods, comprehensive
comparisons of multiple programs \cite{Hall2005,Kuhner1994}, and
detailed testing of the limits of the neighbor-joining algorithm
\cite{Kumar2000}. These studies have concluded that
neighbor-joining is effective for many problems, and have
recommended the algorithm. For example, in \cite{StJohn2003} it is
remarked that ``quartet-based methods are much less accurate
than the simple and efficient method of neighbor-joining''. In a
recent study, Tamura et al. \cite{Tamura2004} conclude that there
are ``bright prospects for the application of the NJ and related methods in inferring large phylogenies.''

Furthermore, new methods are now almost always compared with
neighbor-joining to establish an improvement in performance
\cite{Bruno2000,Farris1996,Gascuel1997,Olsen1994,Ota2000,Ranwez2002,Strimmer1996}.
In other words, neighbor-joining has become the standard by which
new phylogenetic algorithms are compared, and continues to
surface as an effective candidate method for constructing large
phylogenies. This is remarkable, considering the simplicity of the
neighbor-joining algorithm. We begin with a review of the
neighbor-joining algorithm and a very brief description of basic
concepts involved.   
The definitions and notation are based on
\cite{Semple2003}. The reader unfamiliar with the concepts described
below should consult this reference for full details.

Throughout the text a \emph{phylogenetic $X$-tree} $T$ denotes a binary
tree describing the evolutionary relationships between a set of taxa
$X$, which label the leaves of the tree. We will also use the term
\emph{phylogenetic tree} when the set $X$ is clear from the context. A
\emph{cherry} of $T$ is a pair of leaves (or taxa) $(i,j)\in { X \choose
  2}$ such that the path from $i$ to $j$ in $T$ has length exactly 2
(i.e. $i,j$ have a common ``parent'' in $T$). 

A split $A|B$ of $X$ is a bipartition $X=A\cup B$ and $A\cap
B=\emptyset$. In general we will use splits of
$X$ induced by removing an internal edge of $T$ (which will result in
creating two disconnected trees, one with leaf set $A$ and the other
with leaf set $B$). 

A dissimilarity map on $X$ is a map $\delta:X \times X \rightarrow
\mathbb{R}$ satisfying $\delta(x,y) = \delta(y,x)$, $\delta(x,y) \geq 0$
and $\delta(x,x)=0$ for all $x,y \in X$. A tree metric, or an additive dissimilarity map, is a
dissimilarity map $\delta$ for which there exists a tree $T$ with edge
lengths $l:E(T)\rightarrow (0,\infty)$ such that $\delta(i,j)=\sum_{e\in
  P(i,j)}l(e)$ where $P(i,j)$ is the set of edges in $E(T)$ on the path from $i$
to $j$ in $T$. We note that for an additive dissimilarity map, the tree
topology and edge lengths $l$ are uniquely defined. 
\\

We are now ready to give the full description of neighbor-joining: 
\begin{enumerate}
\item  Given a set of taxa $X$ and a dissimilarity map $\delta:X \times
  X \rightarrow
  {\mathbb R}$ (this is a map that satisfies $\delta(i,j)=\delta(j,i)$
  and $\delta(i,i)=0$), compute the {\em Q-criterion} for
$\delta$
\[
 Q_{\delta}(i,j)= \delta(i,j) - \frac{1}{n-2}\left( \sum_{k \neq i}
 \delta(i,k)
 +\sum_{k \neq j} \delta(j,k) \right).
\]
Then select a pair $a,b$ that minimize $Q_{\delta}$ as motivated
by the following theorem:
\begin{thm}[Saitou-Nei \cite{Saitou1987} and Studier-Keppler
    \cite{Studier1988}]
\label{thm:Saitou} Let $\delta_T$ be the tree metric corresponding
to the tree $T$. The pair $a,b$ that minimizes $Q_{\delta_T}(i,j)$
is a cherry in the tree.
\end{thm}
\item If there are more than three taxa, replace the putative
cherry $a$ and $b$ with a leaf $j_{ab}$, and construct a new
dissimilarity map where $\delta(i,j_{ab}) =
\frac{1}{2}(\delta(i,a)+\delta(i,b))$. This is called the {\em
reduction step}. \item Repeat until there are three taxa.

\end{enumerate}

Although the $Q$-criterion is easy to compute, the formula seems,
at first glance, somewhat contrived and mysterious. However, the
formulation of the $Q$-criterion is not accidental and has
many useful properties. For example, it is linear in the
distances, is permutation equivariant (the input order of the taxa
don't matter), and it is {\em consistent}, i.e., it correctly
finds the tree corresponding to a tree metric. Bryant
\cite{Bryant2005} has shown that the $Q$-criterion is in
fact the unique selection criterion satisfying these properties.
Gascuel and Steel \cite{Gascuel2006} in an excellent review, provide a
very precise mathematical
answer to ``what does neighbor-joining do?'', via
a proof that neighbor-joining is a greedy algorithm which ``decreases
the whole tree
length'' as computed by Pauplin's formula \cite{Desper2005,Pauplin2000}.

These results motivate the neighbor-joining algorithm, but they do
not offer any insight into its performance with dissimilarity maps
that are not tree metrics. More importantly, they do not address
the central question of the behavior of neighbor-joining on
dissimilarity maps that arise from maximum likelihood estimation
of distances between sequences in multiple alignments. There is
one result that addresses precisely these issues:

\begin{thm}[Atteson \cite{Atteson1999}]
Neighbor-joining has $l_{\infty}$ radius $\frac{1}{2}$.
\end{thm}

This means that if the distance estimates are at most half the
minimal edge length of the tree away from their true value then
neighbor-joining will reconstruct the correct tree. Atteson's
theorem shows that neighbor-joining is {\em statistically
consistent}. Informally, this means that neighbor-joining
reconstructs the correct tree from dissimilarity maps estimated
from sufficiently long multiple alignments. This has been a widely
used justification for the observed success of neighbor-joining, and is
regarded as the definitive explanation for ``when does neighbor-joining
work?'' \cite{Gascuel2006}.

However, as noted in \cite{Levy2006}, Atteson's condition
frequently fails to be satisfied even when neighbor-joining is
successful. This is also remarked on in \cite{Elias2005}: ``In
practice, most distances are far from being nearly additive [satisfying
  Atteson's condition]. Thus,
although important, optimal reconstruction radius is not
sufficient for an algorithm to be useful in practice.'' Our main
result is an explanation of why neighbor-joining
is useful in practice. We obtain our results using a new consistency
theorem (Theorem \ref{thm:main} in Section 4). Roughly speaking,
our theorem states that neighbor-joining is successful (globally) when
it
works correctly (locally) for the quartets in the tree. Thus, Theorem
\ref{thm:main} provides a crucial link between neighbor-joining
and quartet methods, a connection that is first developed in
Section 3. We also show that Atteson's theorem is a special case
of our theorem.

In Section 5 we present a proof of Atteson's conjecture on the
optimal edge radius of neighbor-joining.
For a dissimilarity map
$\delta$ whose $l_\infty$ distance to a tree metric $\delta_T$ is
less than $\frac{\epsilon}{4}$, we prove that the output $T'$ of
neighbor-joining applied to $\delta$ will contain all edges of $T$
of length at least $\epsilon$, i.e., neighbor-joining has
$l_\infty$ edge radius $\frac{1}{4}$. We say that $T'$
contains the edge $e$ if there exists an edge $e' \in T'$
such that the split of the taxa induced by removing $e'$ from $T'$ is
the same as that
induced by removing $e$ from $T$. This result is tight, as
 \cite{Atteson1999} provides a counterexample for edge radius
 larger than $\frac{1}{4}$. The methods we employ for this result are
 virtually the same as the ones used in the proof of Theorem 16,
 although the proof is non-constructive and based on an averaging
 argument. We conclude with a simulation study that establishes the
 practical significance of our theorems in Section 6.

\section{The shifting lemma}
In this section we provide a few preliminary observations that
are necessary for our results. We begin by reviewing
a lemma from \cite{Gascuel1994}, namely
that there is an alternative to the reduction step (2) in the neighbor
joining algorithm. Condition (2'): if
there are more than three taxa, replace the putative cherry $a$
and $b$ with a leaf $j_{ab}$, and construct a new dissimilarity
map where $\delta(i,j_{ab}) =
\frac{1}{2}(\delta(i,a)+\delta(i,b)-\delta(a,b))$.

Notice that the only difference between (2) and (2')
consists of
addition of the $-\delta(a,b)$ term. On a tree metric, the
first version of the algorithm corresponds to replacing a cherry
by a single node whose distance to the rest of the tree is the
average of the distances of the two collapsed nodes. In the second
version, the reduction step is equivalent with replacing the
cherry by its ``root'', i.e., the point where the path between the
cherry nodes connects to the rest of the tree.

\begin{defn}
We say the dissimilarity map $\delta'$ is a \emph{shift} of $\delta$
if and only if there exists a fixed $\epsilon$ and a distinguished
taxon $a$, such that $\delta'(a,x)=\delta(a,x)+\epsilon$ for all
$x\neq a$ and $\delta'(x,y)=\delta(x,y)$ for all $x,y\neq a$.
\end{defn}

\begin{lemma}[The shifting lemma \cite{Gascuel1994}]
  \label{lem:shifting}
Shifting does not affect the outcome of neighbor-joining.
\end{lemma}

{\bf Proof}: This follows from the observation that shifting by
$\epsilon$ around $a$ changes the value of $Q(x,y)$ by exactly
$-2\epsilon$, for all pairs $x,y$. Moreover, the result of
collapsing taxa $x,y$ in the shifted dissimilarity map is the same as
the
result of collapsing the same taxa in the initial dissimilarity map, or
an
$\epsilon$ shift of it. By these two observations, at each step
neighbor-joining
will collapse the same pairs as before the shift. \qed

\begin{cor}
The two versions of neighbor-joining are equivalent: they
produce trees with the same topology.
\end{cor}
{\bf Proof}: Collapsing $x,y$ by the second reduction method gives
a dissimilarity map that is a $\frac{\delta(x,y)}{2}$-shift of the one
produced by collapsing by the first type of reduction step. \qed

It is important to notice that the operation of shifting a real
tree metric $\delta_T$ by $\epsilon$ around taxon $a$ corresponds
exactly to modifying the length of the leaf edge of $T$
corresponding to $a$ by exactly $\epsilon$. So in effect we can
allow negative leaf edges since shifting around all leaves by a
large enough constant will make them positive, while the outcome
of the algorithm is the same. Of course, the statement of our edge
radius results does not make sense in the case of negative edge
lengths. However, the only negative edges are the leaf edges, and
in that case the statements we make are vacuous. They hold
trivially since neighbor-joining reconstructs the bipartition consisting
of one
taxon vs. the rest of the taxa set correctly, regardless of the
input.

In the remainder of the paper, by a tree metric we will mean a
shift of a tree metric, i.e. a metric corresponding to a tree
where leaf edges are allowed to be negative.

\section{Quartets and neighbor-joining}

We now show that for four taxa, neighbor
joining is equivalent to the {\em four point method}
\cite{Erdos1999}. We will use the notation $(ij:kl)$ to denote the tree
topology on the set $\{i,j,k,l\}$ where the pairs $(i,j)$ and $(k,l)$
form cherries separated by a middle edge. When $i,j,k,l$ are leaves
(internal or external) of a tree $T$, we will say that $(ij:kl)$ is a
quartet of $T$ if the topology induced by $T$ on the four nodes is that
indicated by $(ij:kl)$ .

\begin{prop}
\label{prop:fourtaxa} Let $X=\{i,j,k,l\}$ and $\delta:X \times X
\rightarrow {\mathbb R}$ be a dissimilarity map. The
neighbor-joining algorithm will return the tree $(ij:kl)$ where
$\delta(i,j)+\delta(k,l) \leq min(\delta(i,k) +
\delta(j,l),\delta(i,l)+\delta(j,k))$.
\end{prop}

This result can be easily derived using the $Q$-criterion, but we
prefer to motivate it using an alternative formulation of the
neighbor-joining criterion formulated in \cite{Gascuel1994}.

For a dissimilarity map $\delta$, let
\[ w_{\delta}(ij:kl) = \frac{1}{2} \left(
\delta(i,k)+\delta(i,l)+\delta(j,k)+\delta(j,l) \right)
-\delta(i,j)-\delta(k,l). \] Note that for a quartet $(ij:kl)$ in
a tree $T$ with corresponding tree metric $\delta_T$,
$w_{\delta_T}(ij:kl)$ is double the length of the internal edge in
the quartet. In \cite{Gascuel1994} this is called a ``neighborliness''
measurement.

\begin{thm}
\label{thm:Zcriterion} If $\delta_T$ is the tree metric
corresponding to a phylogenetic $X$-tree $T$ and
\[ Z_{\delta_T}(i,j) = \frac{1}{{n-1 \choose 2}}\sum_{(k,l) \in {{X
      \backslash
      \{i,j\}}\choose{2}}} w_{\delta_T}(ij:kl) \]
then the pair $a,b$ that maximizes $Z_{\delta_T}(i,j)$ is a cherry
in the tree.
\end{thm}

{\bf Proof}: Let $n=|X|$. The sum in $Z_{\delta_T}$ is over unordered
pairs of $X-\{i,j\}$. Observe that for any $\delta$
\[ Z_{\delta}(i,j) = -L_{\delta}(T) - Q_{\delta}(i,j) \]
where $L_{\delta}(T) = \frac{1}{{n-1 \choose 2}}\sum_{x,y \in T}
\delta_T(x,y)$ does not depend
on $i$
or $j$. The theorem now follows directly from Theorem
\ref{thm:Saitou}. \qed

Although the naive computation of the {\em $Z$-criterion} requires
quadratic time, the equivalence with the $Q$-criterion shows that
each entry in the $Z$-matrix is just a sum of a linear number of
distances. One may therefore wonder why the $Z$-criterion is worth
mentioning at all. We outline a number of reasons why the
$Z$-criterion may be a more natural way to formulate the
neighbor-joining selection criterion. For example, note that in
the case of four taxa $i,j,k,l$, the $Z$-criterion is just
$Z_{\delta}(i,j)=\frac{1}{3}w_{\delta}(ij:kl)$ and Proposition
\ref{prop:fourtaxa} follows immediately from Theorem
\ref{thm:Zcriterion}. The $Z$-criterion also highlights the fact
that for a tree metric, the neighbor-joining selection criterion
does not depend on the length of edges adjacent to leaves. This is
remarked on in the proof of the consistency of neighbor-joining in
\cite{Bryant2005}. Furthermore, for a dissimilarity map $\delta$,
$Z_{\delta}(i,j)$ is precisely the difference between the balanced
minimum
length of $\delta$ with respect to the star tree, and the length with
respect to the tree containing the cherry $(i,j)$ with
the remaining taxa unresolved (see Figure 1 in \cite{Gascuel2006} and
the accompanying discussion).

Finally, the $Z$-criterion highlights
the connection between neighbor-joining and quartet methods.
Recall that the naive quartet method consists of choosing a
quartet for each four taxa using the four point method
(Proposition \ref{prop:fourtaxa}), and then returning the tree
consistent with all the quartets (if such a tree exists). This
leads us to
\begin{defn}
\label{defn:quartetconsistency} A dissimilarity map $\delta$ is
{\em quartet consistent} with a tree $T$ if for every $(ij:kl) \in
T$, $w_{\delta}(ij:kl)>max(w_{\delta}(ik:jl),w_{\delta}(il:jk))$.
\end{defn}
By definition, the naive quartet method will reconstruct a tree $T$ from
a
dissimilarity map $\delta$ if $\delta$ is quartet consistent with
$T$. We note that this is essentially the same as the ADDTREE method
\cite{Sattath1977}, with the minor difference that ADDTREE always
outputs a tree,
albeit potentially the wrong one if $\delta$ is not quartet consistent
with $T$.

In the next section we will prove the following extension of
Proposition \ref{prop:fourtaxa}:

\begin{thm}
\label{thm:fourtoseven} If $4 \leq |X| \leq 7$ and $\delta:X
\times X \rightarrow {\mathbb R}$ is a dissimilarity map that is quartet
consistent with a binary tree $T$ then the neighbor-joining
algorithm applied to $\delta$ will construct a tree with the same
topology as $T$. Furthermore, if $5 \leq |X| \leq 7$ then there
exists $\epsilon > 0$ such that if $||\tilde{\delta} - \delta
||_{\infty} < \epsilon$, neighbor-joining applied to
$\tilde{\delta}$ will reconstruct a tree with the same topology as
$T$.
\end{thm}

As we have pointed out, neighbor-joining is equivalent to the
naive quartet method and ADDTREE for $|X|=4$. Theorem
\ref{thm:fourtoseven}
states that neighbor-joining is at least as good as the naive
quartet method for trees with at most $7$ taxa, and is in fact
robust to small changes in the metric.
\begin{figure}[ht]
  \begin{center}
   \includegraphics[scale=0.7]{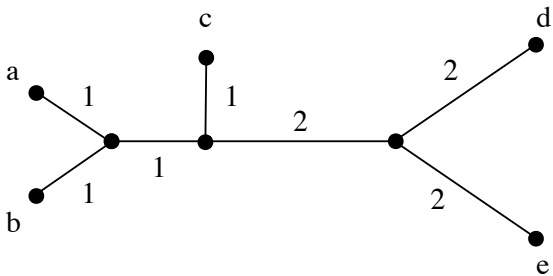}
\caption{A five leaf tree.}
  \end{center}
\end{figure}
\begin{example}
Let $T$ be the 5 leaf tree shown in Figure 1 that corresponds to
the tree metric $\delta_T$, and consider the distorted
dissimilarity map $\delta$
\[ \delta_T \quad = \quad \bordermatrix{ & a & b & c & d & e \cr
a & 0 & 2 & 3 & 6 & 6\cr b & 2 & 0 & 3 & 6 & 6\cr c & 3 & 3 & 0 &
5 & 5\cr d & 6 & 6 & 5 & 0 & 4\cr e & 6 & 6 & 5 & 4 & 0\cr },
\qquad \delta \quad = \quad \bordermatrix{ & a & b & c & d & e \cr
a & 0 & 2 & 3 & 6 & 3\cr b & 2 & 0 & 3 & 6 & 6\cr c & 3 & 3 & 0 &
5 & 5\cr d & 6 & 6 & 5 & 0 & 4\cr e & 3 & 6 & 5 & 4 & 0\cr }.
\]
Note that $\delta$ is not quartet consistent with $T$, because
$\delta(a,e)+\delta(b,c) < \delta(a,b)+\delta(c,e)$. However it is
easy to verify that neighbor-joining constructs a tree with the
same topology as $T$. The example shows that neighbor-joining can
construct the correct tree even when the naive quartet method and
ADDTREE
fail.
\end{example}
The next example shows that Theorem \ref{thm:fourtoseven} fails
for trees with more than $7$ taxa.
\begin{example}
Let $T$ be the 8 leaf tree shown in Figure 2 and let $\delta_T$ be its
corresponding tree metric.
\begin{figure}[ht]
  \begin{center}
   \includegraphics[scale=0.7]{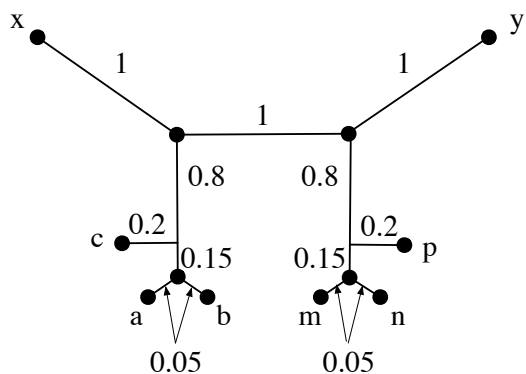}
\caption{An eight leaf tree.}
  \end{center}
\end{figure}
\[ \delta_T \quad = \quad \bordermatrix{ & x & y & a & b & c & m & n & p
  \cr
x & 0 & 3 & 2 & 2 & 2 & 3 & 3 & 3\cr y & 3 & 0 & 3 & 3 & 3 & 2 & 2
& 2\cr a & 2 & 3 & 0 & 0.1 & 0.4 & 3 & 3 & 3\cr b & 2 & 3 & 0.1 &
0 & 0.4 & 3 & 3 & 3\cr c & 2 & 3 & 0.4 & 0.4 & 0 & 3 & 3 & 3\cr m
& 3 & 2 & 3 & 3 & 3 & 0 & 0.1 & 0.4\cr n & 3 & 2 & 3 & 3 & 3 & 0.1
& 0 & 0.4\cr p & 3 & 2 & 3 & 3 & 3 & 0.4 & 0.4 & 0\cr },\]
Consider the
distorted dissimilarity map
\[\delta \quad = \quad \bordermatrix{& x & y & a & b & c & m & n & p
\cr x & 0 & 2.7 & 2.6 & 2.6 & 2.6 & 4.4 & 4.4 & 4.4\cr y & 2.7 & 0
& 4.4 & 4.4 & 4.4 & 2.6 & 2.6 & 2.6\cr a & 2.6 & 4.4 & 0 & 0.1 &
0.4 & 2.7 & 2.7 & 2.7\cr b & 2.6 & 4.4 & 0.1 & 0 & 0.4 & 2.7 & 2.7
& 2.7\cr c & 2.6 & 4.4 & 0.4 & 0.4 & 0 & 2.7 & 2.7 & 2.7\cr m &
4.4 & 2.6 & 2.7 & 2.7 & 2.7 & 0 & 0.1 & 0.4\cr n & 4.4 & 2.6 & 2.7
& 2.7 & 2.7 & 0.1 & 0 & 0.4\cr p & 4.4 & 2.6 & 2.7 & 2.7 & 2.7 &
0.4 & 0.4 & 0\cr }
\]
 that is quartet consistent with $T$.
It is easy to see that $Q_{\delta}(x,y)=-6.24$, while
$Q_{\delta}(a,b)=Q_{\delta}(m,n)=-6.04$. Therefore the function
$Q_{\delta}$ is not minimized at one of the cherries of $T$,
and the neighbor-joining algorithm applied to $\delta$ outputs a tree
different from $T$, in which $x,y$ form a cherry.
\end{example}

Fortunately, there is a single extra condition which ensures
 that neighbor-joining correctly reconstructs a tree. In what follows
 we say that  a leaf $x$ is {\em interior to a
 quartet} $(ij:kl)$ in a tree $T$ if none of $(ik:xl)$, $(ik:xj)$,
 $(ix:jl)$, or $(kx:jl)$ are quartets in $T$.
\begin{figure}[ht]
\label{fig:quartetadditivity}
  \begin{center}
   \includegraphics[scale=0.7]{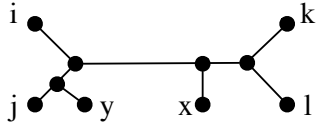}
\caption{The quartet additivity configuration.}
  \end{center}
\end{figure}
\begin{defn}
\label{def:quartetadditivity} A dissimilarity map $\delta:X \times
X \rightarrow {\mathbb R}$ is \emph{quartet additive} with a tree
$T$ if for every $(ij:kl) \in T$ with $x$ interior to $(ij:kl)$,
and $y$ not interior to $(ij:kl)$ such that $(ij:xy)$ is not a
quartet of $T$, we have $w(kl:xy)>w(ij:xy)$ (see Figure 3).
\end{defn}
We conclude with three basic lemmas that are important in the next
section. The proofs are left as an exercise for the reader.

\begin{lemma} Quartet consistency and additivity are both invariant with
  respect to
the shifting operation.
\end{lemma}

\begin{lemma}[Spectator Lemma]
\label{spectator}
For any $a, b, x, y, t \in X$,
\[  2 w(ab:xy)  =   w(tb:xy) + w(at:xy) + w(ab:ty) + w(ab: xt). \]
\end{lemma}
The Spectator Lemma is used to prove
\begin{lemma}
\label{spectator1}
For any $a, b, c, i, j, x, y \in X$,
\begin{eqnarray*}
 3w(ab:xy) - 3w(ac:xy) & = & w(ab:xc) - w(ac:xb) + w(ab:yc) -
 w(ac:yb),\\
 4w(ab:xy) - 4w(ij:xy) &  = & w^*(ab:x:ij) + w^*(ab:y:ij),
\end{eqnarray*}
{\rm where} $w^*(ab:x:ij) = w(ab:ix) + w(ab:jx) - w(ax:ij) - w(bx:ij)$.
\end{lemma}

\section{A consistency theorem for neighbor-joining}

\begin{thm}
\label{thm:main} If $\delta:X \times X \rightarrow {\mathbb R}$ is
quartet consistent and quartet additive with a tree $T$, then
neighbor-joining applied to $\delta$ will construct a tree with
the same topology as $T$.
\end{thm}

The proof of the theorem consists of two main parts. First, we show that
when the reduction step collapses a pair of taxa $(x,y)$ which form a
cherry in $T$, then the dissimilarity metric given by reducing $\delta$
is quartet consistent and additive with the tree $T'$ obtained by
clipping off the cherry $(x,y)$ and labeling its former root (now a
leaf) with the new taxon thus created in the reduction step. In other
words, the node obtained by collapsing $(x,y)$ in
$\delta$ is assigned to the location of the common ancestor of $x$ and
$y$ in the true tree $T$. Note that this only makes sense when $x$ and
$y$ do indeed form a cherry, as this is the only case in which their
common ancestor is well defined. The second part of the inductive proof
is showing that given $\delta$ quartet consistent and quartet additive
with a tree $T$, the optimal pair for the reduction step is guaranteed
to form a cherry in $T$. This clearly completes the inductive
argument. We proceed with the first step.

\begin{lemma}
Quartet consistency is maintained when reducing a cherry of the
reference tree. Formally, given a pair of taxa $x,y\in X$ which form a
cherry in $T$, the result of collapsing $(x,y)$ in $\delta$ is quartet
consistent with the topology given by deleting the leaf edges leading to
$x$ and $y$ in $T$ and assigning the new member of the set of taxa to
the new leaf thus formed in $T$.
\end{lemma}

{\bf Proof}: Without loss of generality, we may assume that we are
collapsing taxa $x,y$ into taxon $z$ by using the first type of
reduction step. For a set $U\subset X$, we let $T|_U$ be the topology
induced by $T$ on $U$. Note that for $X_1=X-\{x\}$ and $X_2=X-\{y\}$,
the two topologies $T|_{X_1}$ and $T|_{X_2}$ are isomorphic under the map
$f:X_1\rightarrow X_2$ defined by $f(u)=u$ for $u\neq x,y$ and
$f(y)=x$. We furthermore notice that $\delta|_{X_i}$ is quartet consistent
with $T|_{X_i}$ for $i=1,2$ since $\delta$ is quartet consistent with
$T$. Therefore $\delta|_{X_i}$ will be quartet consistent with $T'$ under
identifying $y$ or $x$ with $z$. We note again that this property holds
\emph{if and only if} $(x,y)$ form a
cherry in $T$.

Now note that the reduced distance metric $\delta'$ on the set
$X'=X-\{x,y\} \cup\{z\}$ is in fact a linear combination of the two
dissimilarity maps $\delta|_{X_1}$ and $\delta|_{X_2}$ under identifying $z$
with $y$ in $X_1$ and $x$ in $X_2$ (We define the restriction of the
dissimilarity map to a subset of its domain in the obvious way). 

The last piece of the proof involves noticing that the condition of
quartet
consistency with respect to a topology $T$ is a set of linear
inequalities (defined by $T$), on the values $\delta(i,j)$, or
``linear in $\delta$'' for short. Concretely, this means that any linear
combination of dissimilarity maps that are quartet consistent with a
topology $T$ will also be quartet consistent with $T$. As noted above,
both $\delta|_{X_1}$ and $\delta|_{X_2}$ are quartet consistent with $T'$,
while $\delta'$ is a linear combination of the two under the obvious
isomorphisms (in fact $\delta'$ is the mean of $\delta|_{X_1}$ and
$\delta|_{X_2}$). We conclude that $\delta'$ is quartet consistent with
$T'$.
\qed\\

\begin{lemma}
Quartet additivity is maintained when reducing a cherry of the reference
tree. Formally, given a pair of taxa $x,y\in X$ which form a cherry in
$T$, the result of collapsing $(x,y)$ in $\delta$ is quartet additive
with respect to the topology given by deleting the leaf edges leading to
$x$ and $y$ in $T$ and assigning the new member of the set of taxa to
the new leaf thus formed in $T$.
\end{lemma}

{\bf Proof}: We note that quartet additivity is also a property that is
linear in $\delta$. The proof proceeds identically with the previous
lemma.

{\bf Proof of Theorem \ref{thm:main}}: By the above lemmas, it
suffices to prove that at any step, the pair of taxa which
maximize the $Z$-criterion form a cherry. We argue by
contradiction. Let us consider a pair $\delta, T$ such that $(i,j)$ is
a pair of taxa which maximizes $Z_{\delta}$, but does not form
a cherry in $T$. Throughout the proof, and in the remainder of the paper, we multiply $Z_{\delta}$ by ${n-1
  \choose 2}$ to simplify the formulas.

\textit{Case 1}: Suppose $i$ or $j$ are part of a cherry.  Without loss
of generality, assume that leaf $i$ forms a cherry with leaf $k \neq j$
and let $X' = X - \{i,j,k\}$.  Then:

$$Z_{\delta}(i,k) - Z_{\delta}(i,j) = \sum_{\forall x \in X'} w(ik:xj) -
w(ij:xk) +
\sum_{\forall (x,y) \in {X' \choose 2}} w(ik:xy) - w(ij:xy).$$
Applying Lemma \ref{spectator1} to the second summand, we have:
\begin{eqnarray*}
Z_{\delta}(i,k) - Z_{\delta}(i,j) &=& \sum_{\forall x \in X'} w(ik:xj) -
w(ij:xk) \\ && +
\frac{1}{3}\sum_{\forall x,y \in {X' \choose 2}} w(ik:xj) - w(ij:xk) +
w(ik:yj) - w(ij:yk) \\
        &=& \frac{n-1}{3} \sum_{\forall x \in X'} w(ik:xj) - w(ij:xk).
\end{eqnarray*}

Since $(ik:xj)$ is a quartet in $T$ for any $x$, by consistency
$w(ik:xj) - w(ij:xk) > 0$ for all $x$ and therefore $Z_{\delta}(i,k) -
Z_{\delta}(i,j) >
0$, a contradiction.

\begin{figure}[ht]
\label{fig:case2}
  \begin{center}
   \includegraphics[scale=0.7]{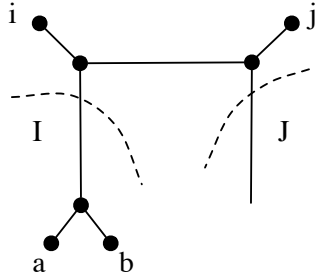}
\caption{Case 2 in Theorem 16.}
  \end{center}
\end{figure}

\textit{Case 2}:
Neither $i$ nor $j$ are part of a cherry.  Then, we
have a situation as in Figure 4.  Let
$I$ be the set of leaves on the subtree nearest $i$ along the path from
$i$ to $j$ and, similarly, let $J$ be the set of leaves on the subtree
nearest $j$.  Without loss of generality, we assume that $|I| \leq |J|$
and let the pair $a,b \in I$ be a cherry in $T$.  Let $X' = X -
\{a,b,i,j\}$. Then
\begin{eqnarray*}
Z_{\delta}(a,b) - Z_{\delta}(i,j) &=& \sum_{x \in X'} w(ab:ix) +
w(ab:jx) - w(ax:ij) -
w(bx:ij) \\
                && + \sum_{x,y \in {X' \choose 2}} w(ab:xy) - w(ij:xy).
\end{eqnarray*}
By Lemma \ref{spectator1},
\begin{eqnarray*}
Z_{\delta}(a,b) - Z_{\delta}(i,j) &=& \sum_{x \in X'} w^*(ab:x:ij) +
                    \frac{1}{4}\sum_{x,y \in {X' \choose 2}}
                    w^*(ab:x:ij) + w^*(ab:y:ij) \\
                &=& \frac{n-1}{4} \sum_{x \in X'} w^*(ab:x:ij)
\end{eqnarray*}

First note that if $x \in X'$ is a leaf that is not in $I$ or in $J$
(i.e., the paths from $i$ to $x$ and $j$ to $x$ intersect the path from
$i$ to $j$), then $w(ab:xy)>w(ij:xy)$ for any leaf $y$ by quartet
consistency. Thus we restrict our attention to the leaves in $I$ and
$J$.
Let $I' = I - \{a, b\}$.  Then, since $|I| \leq |J|$, it follows that
$|I'| \leq |X' - I'|$ and  we choose a subset of $J$ that is the
same size as $I'$.  In particular, there exists $I^* \subset J$
such that $|I^*| = |I'| = P$.  Let $I' = \{x_1, ..., x_P\}$, $I^* =
\{y_1, ..., y_P\}$ and $X'' = X' - I' - I^*$.  Then:

\begin{eqnarray*}
\frac{4}{n-1}\left( Z_{\delta}(a,b) - Z_{\delta}(i,j) \right) &>&
\sum_{p=1}^P
w^*(ab:x_p:ij) + w^*(ab:y_p:ij) +
                        \sum_{x \in X''} w^*(ab:x:ij) \\
                        &=& 4 \sum_{p=1}^P w(ab:x_py_p) - w(ij:x_py_p) +
 w(ij:ax)]  \\ & & + 
             \frac{2}{3}[w(bi:jx) - w(ij:bx)].
\end{eqnarray*}

Since $(ab:ix), (ai:jx)$ and $(bi:jx)$ are all quartets in $T$, by
consistency each term is positive.  Therefore $w^*(ab:x:ij) > 0$ and so
$Z_{\delta}(a,b) > Z_{\delta}(i,j)$, a contradiction. \qed

\begin{remark}
Theorem \ref{thm:fourtoseven} follows from the observation that
quartet consistency suffices in the proof of Theorem
\ref{thm:main} when $4 \leq |X| \leq 7$. Details are omitted. 
\end{remark}

\begin{cor}
\label{cor:main} Neighbor-joining has $l_{\infty}$ radius
of at least $\frac{1}{2}$. Following Atteson's results from
\cite{Atteson1999} for the reverse inequality, we conclude that neighbor-joining has $l_{\infty}$ radius equal $\frac{1}{2}$.
\end{cor}
{\bf Proof}: It is easy to see that if $\delta_T$ is a tree
metric and $\delta$ is a metric with $max_{i,j}|\delta_T(i,j) -
\delta(i,j)| < \frac{1}{2} min_{e \in E(T)}l(e)$ where $l(e)$ is
the length of edge $e$ in $T$, then $\delta$ is quartet consistent
and quartet additive with $T$. \qed

The next corollary extends the Visibility Lemma from
\cite{Elias2005}. A taxon $b$ is {\em visible} from $a$ with
respect to a dissimilarity map $\delta$ if
\[ b = argmax_{x \neq a}Z_{\delta}(a,x). \]

\begin{cor}
\label{cor:visibility} If $\delta$ is quartet consistent with a
tree T and $(a,b)$ is a cherry of $T$, then $b$ is
visible from $a$ with respect to $\delta$.
\end{cor}
{\bf Proof}: This follows directly from the first step of the
proof of Theorem \ref{thm:main}. \qed\\

The Visibility Lemma is the key to developing a fast neighbor
joining algorithm (FNJ) that has optimal run time complexity
$O(n^2)$ \cite{Elias2005}. In fact, we can conclude that
\begin{cor}
\label{cor:FNJ} If $\delta$ is quartet consistent and quartet
additive with respect to a tree $T$ then FNJ will reconstruct $T$
from $\delta$ with run time complexity $O(n^2)$ (which is also the size
of the algorithm input). 
\end{cor}

\section{The Edge Radius of the neighbor-joining Algorithm}
In this section we prove a strengthening of the following conjecture
about the
edge radius of the neighbor-joining algorithm:
\begin{conj}[Atteson \cite{Atteson1999}]
Let $T$ be a phylogenetic $X$-tree with associated tree metric
$\delta_T$, and let
$e$ be an edge of $T$ of length $l(e)$. If $\delta$ is a dissimilarity
map whose $l_\infty$ distance to $\delta_T$ is less than
$\frac{l(e)}{4}$, then neighbor-joining applied to $\delta$ will
reconstruct the edge
$e$ correctly, i.e., the tree $T'$ output by neighbor-joining will
contain an
edge $e'$ which induces the same split in the tree $T'$ as $e$ induces
in $T$.
\end{conj}
Since the necessary $l_{\infty}$ error bound needed for the
neighbor-joining
algorithm to reconstruct an edge correctly is $\frac{1}{4}$ of the
length
of the edge, we say that the edge radius of neighbor-joining is
$\frac{1}{4}$. This result
is optimal. In \cite{Atteson1999}, Atteson presents an example
where the statement fails for $l_{\infty}$ error larger than
$\frac{l(e)}{4}$.

For ease of exposition, in what follows we will drop the
requirement that the input trees are binary. In other words, we will
allow internal nodes of degree higher than three. Note that an internal
node of degree at
least $4$ corresponds to one or more internal edges of length $0$ in a
binary tree. We also continue to allow negative leaf edges. Since in
this section we are only concerned with recovering splits corresponding
to edges of at lest a certain length in the reference tree, edges of
zero size can easily be allowed as our requirements for reconstructing
them become null. Also, since leaf edges are recovered correctly by
design (they correspond to trivial splits), negative
leaf edges can easily be allowed without affecting the analysis. This is
also a consequence of Lemma \ref{lem:shifting}.

\begin{defn}
Let $T$ be a tree and $e$ a non-leaf edge of length $l$ in $T$,
corresponding
to the split $A|B$ of $X$.
We say that a dissimilarity map $\delta$ is \emph{$A|B$-consistent}
 with respect to the tree metric  $\delta_T$ if the following conditions
 hold
\begin{itemize}
\item $\delta(x,y)-\delta_T(x,y)<\frac{l(e)}{4}$ for all pairs $x,y\in
  A$
  and
all pairs $x,y\in B$, \item $|\delta(x,y)-\delta_T(x,y)|<\frac{l(e)}{4}$
  for
all pairs $x\in A$ and $y\in B$.
\end{itemize}
\end{defn}

We are ready to state the main theorem of the section
\begin{thm}\label{thm:edge_radius}
If $\delta$ is an  $A|B$-consistent dissimilarity map with respect to
the tree $T$ and the tree metric $\delta_{T}$, then the neighbor-joining
algorithm applied to $\delta$ will output a tree $T'$ which
contains $A|B$ among its edge-induced splits.
\end{thm}

This implies more than Atteson's conjecture since we
are not imposing a lower bound on the estimated distances between
taxa situated on the same side of the split. This observation is crucial; we
show in Theorem 34 that Atteson's condition as originally stated in
\cite{Atteson1999} does not hold inductively. As we will see, Lemma 7 in
\cite{Atteson1999} fails if one is only trying to recover a specific
edge in the tree, because non-neighboring pairs may be collapsed during
the agglomeration steps. 

The proof of Theorem 25 is based on two propositions. In Proposition 26,
we show that agglomerating
a pair of elements in $A$ or a pair of elements in $B$ preserves
$A|B$-consistency. It therefore suffices
to show that at every step of the algorithm, the $Z$-criterion is
maximized for a pair that lies either
in $A$ or in $B$.  This is shown in Proposition 27 by an averaging
argument. We will assume, by way of contradiction, that there is a
pair of leaves $(i,j)$ with $i \in A$, $j \in B$ and $Z_\delta(i,j)$
maximal.
To obtain a contradiction, we will assume without loss of generality
that $|A| \leq |B|$ and
show that, on average, $Z_\delta(a_1,a_2) - Z_\delta(i,j) > 0$
where the average is taken over all pairs $a_1, a_2 \in {A \choose 2}$.
Consequently, there
must be at least one pair of elements $x, y \in A$ such that
$Z_\delta(x,y) > Z_\delta(i,j)$.

\begin{prop}\label{collapse}
Given an $A|B$-consistent dissimilarity map $\delta$, with respect to a
tree $T$,
collapsing a pair of taxa $x,y\in A$ will result in a metric
$\delta'$ which is $A'|B$-consistent with respect to a tree $T'$.
Here $A'$ is the set of taxa obtained by replacing $x,y$ by the
collapsed node $z$ in $A$.
\end{prop}

\begin{figure}[ht]
  \begin{center}
   \includegraphics[scale=0.65]{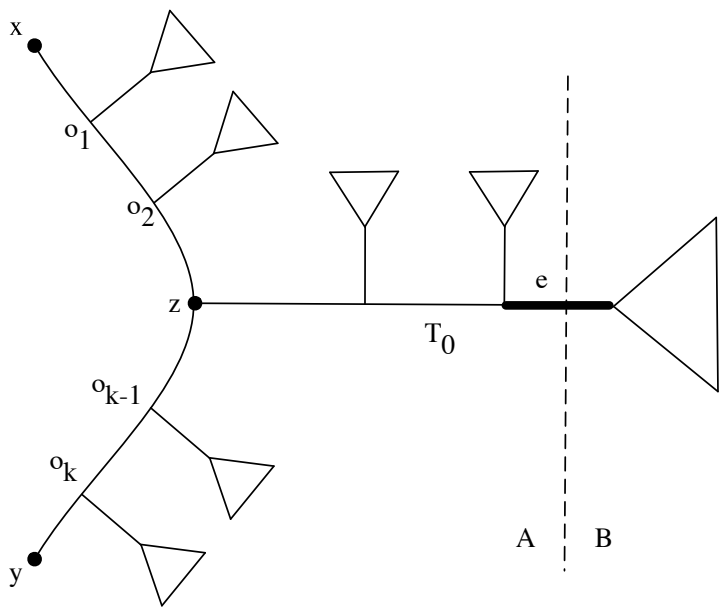}
\caption{The collapsing lemma.}
  \end{center}
\end{figure}

{\bf Proof}: As before, we assume without loss of generality that
we are using the first variant of the reduction step. Now shift
the new dissimilarity map $\delta'$ by $\frac{\delta_{T}(x,y)}{2}$
around
the new taxon $z$. Again, this can be done without affecting the
outcome of the algorithm. In effect, this is equivalent to
defining the distances with respect to $z$ in the following
manner:
$$\delta'(z,a)=\frac{1}{2}(\delta(x,a)+\delta(y,a)-\delta_{T}(x,y)).$$

Now let $e$ be the edge in $T$ corresponding to the $A|B$ split.
Let $l$ be its length. Let $p$ be the path in $T$ that joins $x$
and $y$. Then $e\not\in p$ and let $o$ be the internal point of
$T$ where a path from $e$ reaches $p$. Consider the new tree $T'$
where the taxa $x$ and $y$ are removed and the new taxon $z$ is
placed exactly at the internal point $o$. $\delta_{T'}$ and $A'$
are defined in the obvious way and $\delta'$ is defined by
collapsing $x,y$ in $\delta$ according to the reduction described
above.

Let $T_{0}\ldots T_{k}$ be the subtrees of $T$ hanging off the
path $p$ and let $T_0$ be the one that contains $e$. We now
observe that $\delta'(a,b)=\delta(a,b)$ and
$\delta_{T'}(a,b)=\delta_{T}(a,b)$ for $a,b\neq z$. In this case
the errors remain the same. For $b\in B$, and therefore $b\in
T_{0}$, we observe that
$$\delta_{T'}(z,b)=\delta_{T}(o,b)=\frac{1}{2}(\delta_{T}(x,b)+\delta_{T}(y,b)-\delta_{T}(x,y)).$$
Therefore
$$|\delta'(z,b)-\delta_{T'}(b,z)|=|(\delta(x,b)-\delta_{T}(x,b))+(\delta(y,b)-\delta_{T}(y,b)|\frac{1}{2}<\frac{l}{4}.$$

Now for all $i$ let $o_{i}$ be the point of the path $p$ that is
the root of the subtree $T_{i}$. Then for any $a\in A$, so $a\in
T_{i}$ for $i\neq 0$ or $a\in T_{0}\cap A$,

$$\delta'(z,a)=\delta_{T'}(a,o_{i})+[(\delta(x,a)-\delta_{T}(x,a))+(\delta(y,a)-\delta_{T}(y,a))]/2=$$
$$\delta_{T'}(a,z)+[(\delta(x,a)-\delta_{T}(x,a))+(\delta(y,a)-\delta_{T}(y,a))]/2-\delta_{T}(o,o_{i})<$$
$$\delta_{T'}(a,z)+\frac{l}{4}-\delta_{T}(o,o_{i}).$$

Therefore $\delta'$ is $A'|B$-consistent with respect to $T'$ and
$T'$ ``contains'' the edge $e$.\qed

Now assume that the pair $(i,j)$ which maximizes $Z_\delta(i,j)$ is such
that $i \in A, j \in B$.  Without loss of
generality, we assume that $|A| \leq |B|$.

\begin{prop}
$$S(\delta:A,i,j) = \sum_{(a_1, a_2) \in {A \choose 2}} Z_\delta(a_1,
  a_2) -
Z_\delta(i,j) > 0.$$
That is, there exist $x,y \in {A \choose 2}$ such that $Z_{\delta}(x,y)>
Z_{\delta}(i,j)$.
\end{prop}

We prove the proposition by comparing the differences between the
dissimilarity values and the true tree metric $\delta_T$, with counts of
the contested edge in the averaging step. The calculation is elementary
but tedious, and requires a few lemmas and some notation:

Let $M$ be the set of all dissimilarity maps $\delta:X \times X
\rightarrow {\mathbb R}$ on a set $X$. We represent a linear function
$f:M
\rightarrow {\mathbb R}$ by
$$\delta \mapsto \sum_{\forall (a,b) \in {X \choose 2}} \alpha_f(a,b)
\delta(a,b).$$
Similarly, if $M_{\tau}$ is the set of all tree metrics on a set $X$, we
represent a
linear function $f:M_{\tau} \rightarrow {\mathbb R}$ by
$$\delta_T \mapsto \sum_{\forall e \in E(T)} \beta_{f,T}(e) l(e).$$
More explicitly, $\beta_{f,T}(e)$ is the coefficient of $l(e)$ in
$f(\delta_T)$, when the metric $\delta_T$ corresponds to the edge
lengths $l$.
  If $T$ is obvious from the context, we will write $\beta_{f,T}$ as
  $\beta_f$.

For a tree $T$, note that an edge $e$ divides the leaves into two sets.
For a given tree leaf $i$, we define $N^+_e(i)$ as the set of leaves
on
the same side of $e$ as $i$ and $N^-_e(i)$ as the set of leaves on the
far
side of $e$ from $i$.  By a slight abuse of notation, we will also use
$N^+_e(i)$ and $N^-_e(i)$ to denote the number of elements in their
respective sets when such use does not give rise to confusion.

\begin{lemma}\label{beta}
Let $a,b$ be leaves in a tree $T$.  Then

$$
\beta_{Z(a,b)}(e) = \left\{
\begin{array}{ll}
- (N^+_e(a) - 1)(N^-_e(a) - 1)
       & \mbox{ if } e \in P_{ab}; \\
       & \\
N^-_e(a)(N^-_e(a) -1),
       & \mbox{otherwise}.
\end{array}
\right.
$$
\end{lemma}
{\bf Proof}: For a tree metric $\delta_T$, $w_{\delta_T}(ab:xy)$ is
twice the length of the splitting path if $(ab:xy)$ is
a quartet of $T$ and negative the length of the path otherwise.  Hence,
if an edge $e$ is on the path from $a$ to $b$, then
there is no quartet $(ab:xy)$ such that $e$ is on the splitting path.
Hence, the edge $e$ is only counted negatively, once for every element
$x \in N^+_e(a) - \{a\}$ and $y \in N^+_e(b) - \{b\} = N^-_e(a) -
\{b\}$.  There are exactly $(N^+_e(a)-1)(N^-_e(a)-1)$ such pairs
$(x,y)$.

If $e$ is not on the path from $a$ to $b$, then for any pair of elements
$(x,y) \in {N^-_e(a) \choose 2}$, $(ab:xy)$ are exactly the quartets of
$T$ of the form $(ab:\cdot\cdot)$ with $e$ is on
the splitting path.  Consequently, $\beta_{Z(a,b)}(e) = 2{N^-_e(a)
  \choose 2} = N^-_e(a)(N^-_e(a) -1)$. \qed


\begin{lemma}\label{lowerbound}
Let an edge $e$ define a split $A|B$ and assume $|A| \leq |B|$. Then
\[ S(\delta_T:A,i,j) \geq {|A| \choose 2}(|B| - 1)(n - 1)l(e).\]
\end{lemma}

{\bf Proof}: This is equivalent to showing that
\[ \beta_{S(\delta_T:A,i,j)}(e) = {|A| \choose 2}(|B| - 1)(n - 1),\]
and for any other edge $e' \in T$, $\beta_{S(\delta_T)}(e') \geq 0$.
Note that since $e$ is never on the path between $a_1$ and $a_2$ for any
pair $a_1, a_2 \in A$, $\beta_{Z(a_1, a_2)}$
is the same for any choice of $a_1, a_2$.  Also, $e$ is on the
path between $i$ and $j$, so by application of Lemma \ref{beta}:
\begin{eqnarray*}
\beta_{S(\delta_T:A,i,j)}(e) &=& \sum_{(a_1, a_2) \in {A \choose 2}}
\beta_{Z(a_1, a_2)}(e) - \beta_{Z(i,j)}(e) \\
                       &=& {|A| \choose 2}[\beta_{Z(a_1, a_2)}(e) -
  \beta_{Z(i,j)}(e)] \\
                       &=& {|A| \choose 2}[|B|(|B| - 1) + (|A| - 1)(|B|
  - 1)] \\
                       &=& {|A| \choose 2}(|B| - 1)(n - 1). \\
\end{eqnarray*}

Let $t$ be either vertex of the edge $e$.
Assume we have an edge $e' \neq e$ such that $e'$ is in the subtree
spanned by $B$, denoted $e' \in [B]$.  Then $e'$
is never on the path between $a_1$ and $a_2$, so $\beta_{Z(a_1,
  a_2)}(e') =
N^-_{e'}(t)(N^-_{e'}(t) - 1)$ for any pair $a_1, a_2 \in A$.  If $e'$ is
on the path between $i$ and $j$, then $\beta_{Z(i, j)}(e') < 0$, so
clearly
$\beta_{S(\delta_T:A,i,j)}(e') > 0$.  If $e'$ is not on the path between
$i$
and $j$, then
$\beta_{Z(i, j)}(e') = \beta_{Z(a_1, a_2)}(e')$ and
$\beta_{S(\delta_T)}(e') = 0$. 
The proof of the final case that
needs to be considered, namely $e' \in [A]$ consists of a trivial, yet
slightly lengthy, counting argument. We only present a brief sketch. 

Again, in the case when $e'$ is on the path between $i$ and $j$, then
$\beta_{Z(i, j)}(e') < 0$, so
clearly $\beta_{S(\delta_T:A,i,j)}(e') > 0$. For $e'$ not on the path
from $i$ to $j$, let $A'=N_{e'}^{-}(i)\subset A$. Let
$\alpha=|A'|$. Then a simple counting argument shows that $\beta_{Z(i,
  j)}(e')=2{\alpha \choose 2 }$, whereas for $f=\sum_{(a_1, a_2) \in {A
    \choose 2}} Z_\delta(a_1, a_2)$ we have $\beta_{f}(e')=2{\alpha
  \choose 2 }{|B| \choose 2}$. Since $|B|>|A|$, this concludes the
proof. 
\qed

\begin{lemma}\label{alpha}
Let $a,b$ be elements of the leaf set $X$ of size $n$.  Then:
$$
\alpha_{Z(a,b)}(x,y) = \left\{
\begin{array}{ll}
- {n-2 \choose 2}
       & \mbox{ if } |\{a,b\}\cap\{x,y\}| = 2; \\
       & \\
\frac{1}{2}(n-3)
       & \mbox{ if } |\{a,b\}\cap\{x,y\}| = 1; \\
       & \\
-1
       & \mbox{ if } |\{a,b\}\cap\{x,y\}| = 0.
\end{array}
\right.
$$
\end{lemma}
{\bf Proof}: The term $\delta(a,b)$ occurs in all ${n-2 \choose 2}$
$w(ab:xy)$ terms in $Z(a,b)$, each time with a coefficient of $-1$.  To
compute $\alpha_{Z(a,b)}(a,x)$, we note that $\delta(a,x)$ occurs in all
$(n-3)$ terms of the form $w(ab:x\cdot)$ with a coefficient of
$\frac{1}{2}$.  The same holds for $\alpha_{Z(a,b)}(b,x)$.  Lastly,
$\delta(x,y)$ for $x,y \neq a,b$ occurs only in one term, $w(ab:xy)$,
with coefficient $-1$. \qed

\begin{lemma}\label{8cases}
Let $a, a_1 \neq a_2 \in A - \{i\}$ and $b, b_1 \neq b_2 \in B -
\{j\}$. We have
\begin{enumerate}
\item $\alpha_{S(\delta)}(i,j) = \frac{1}{2}(|A|-1)(|B|-1) + {|A|
  \choose 2}{n-2 \choose 2},$ \\
\item $\alpha_{S(\delta)}(i,a) = -{|B| \choose 2} - \frac{1}{2}(n-3){|A|
  \choose 2},$ \\
\item $\alpha_{S(\delta)}(j,a) = \frac{1}{2}(|A|-1)(|B|-1) -
  \frac{1}{2}(n-3){|A| \choose 2} = -\frac{1}{4}(n-1)(|A|-1)(|A|-2),$ \\
\item $\alpha_{S(\delta)}(i,b) = \frac{1}{2}(|A|-1)(|B|-1) -
  \frac{1}{2}(n-3){|A| \choose 2} = -\frac{1}{4}(n-1)(|A|-1)(|A|-2),$ \\
\item $\alpha_{S(\delta)}(j,b) = -{|A| \choose 2} - \frac{1}{2}(n-3){|A|
  \choose 2},$ \\
\item $\alpha_{S(\delta)}(a,b) = \frac{1}{2}(|A|-1)(|B|-1) + {|A|
  \choose 2},$ \\
\item $\alpha_{S(\delta)}(a_1,a_2) = -{|B| \choose 2} + {|A| \choose
  2},$ \\
\item $\alpha_{S(\delta)}(a_1,a_2) = -{|A| \choose 2} + {|A| \choose 2}
  = 0.$ \\
\end{enumerate}
There are $1, |A|-1, |A|-1, |B|-1, |B|-1, (|A|-1)(|B|-1), {|A|-1 \choose
  2}$ and ${|B|-1 \choose 2}$ of each
of these terms, respectively.
\end{lemma}
{\bf Proof}:
Let $S_1(\delta) = \sum_{(a_1, a_2) \in {A \choose 2}} Z_\delta(a_1,
a_2)$  Note that
$$
\alpha_{S_1(\delta)}(x,y) = \left\{
\begin{array}{ll}
- {|B| \choose 2}
       & \mbox{ if } |\{x,y\}\cap A| = 2; \\
       & \\
\frac{1}{2}(|A|-1)(|B|-1)
       & \mbox{ if } |\{x,y\}\cap A| = 1; \\
       & \\
- {|A| \choose 2}
       & \mbox{ if } |\{x,y\}\cap A| = 0.
\end{array}
\right.
$$
 We provide the proof for the first case, where $x, y \in A$ (the other
 cases are similar).  Let $A' = A - \{x,y\}$, then:
\begin{eqnarray*}
\alpha_{S_1(\delta)} &=& \alpha_{Z_\delta(x,y)}(x,y) + \sum_{a \in
  A'}\alpha_{Z_\delta(a,x)}(x,y) + \alpha_{Z_\delta(a,y)}(x,y) +
\sum_{a_1,a_2 \in {A' \choose 2}} \alpha_{Z_\delta(a_1,a_2)}(x,y)\\
                     &=& -{n-2 \choose 2} + |A'|\left[\frac{1}{2}(n-3) +
  \frac{1}{2}(n-3)\right] + {|A'| \choose 2}(-1) \\
                     &=& \frac{1}{2}(n-|A|)\left[|A|-n+1\right] \quad =
  \quad -{|B| \choose 2}.
\end{eqnarray*}
Identities (1)--(8) now follow after some elementary algebra and Lemma
\ref{alpha}.\qed

\begin{lemma}
$$S(\delta_T) - S(\delta)  < \frac{l}{8}(|A|-1)(n-1)(3|A|n - 2n - 2|A|^2
  - 4|A| + 4)$$
\end{lemma}
{\bf Proof}:
Let $\delta$ be an $A|B$-consistent metric and set
$\tilde{\delta} = \delta_T - \delta$. Note that
for all $x,y \in {X \choose 2}$,
\[ \alpha_{S(\tilde{\delta})}(x,y) \tilde{\delta}(x,y) <
|\alpha_{S(\tilde{\delta})}(x,y)|\frac{l}{4}. \]
This follows directly from the fact that $\alpha_{S(\delta)}(x,y) < 0$
for all cases where $x,y \in A$ or $x,y \in B$, together with the signs
of the terms in the Lemma \ref{8cases} and the definition of
$A|B$-consistency. It follows that
\[
S(\tilde{\delta}) = \sum_{x,y \in {X \choose
    2}}\alpha_{S(\tilde{\delta})}(x,y)\tilde{\delta}(x,y)
                  <  \sum_{x,y \in {X \choose
    2}}|\alpha_{S(\tilde{\delta})}(x,y)|\frac{l}{4}, \]
and the lemma follows by summing the terms appearing in Lemma
    \ref{8cases}.\qed

{\bf Proof of Proposition 27 and Theorem 25}: By Lemma 29 and 32 it
suffices to show that
$$\frac{l}{2}|A|(|A|-1)(|B| - 1)(n - 1) - \frac{l}{8}(|A|-1)(n-1)(3|A|n
- 2n - 2|A|^2 - 4|A| + 4) \geq 0.$$
This inequality follows from the fact that we chose $|A| \leq |B|$ and
consequently
$2|A| \leq n$. Thus, the pair of leaves that
maximize $(Z \cdot,\cdot)$ are both in $A$. The theorem now follows from
Proposition 26. \qed

We conclude by stating that our analysis holds trivially for the
fast neighbor-joining algorithm of \cite{Elias2005}. This follows
from the observation that for an $A|B$-consistent dissimilarity map
$\delta$, no
pair $x,y$ with $x\in A$ and $y\in B$ can maximize the
$Z(\cdot,\cdot)$ criterion, and therefore the maximizing pair,
which has to be visible from both of its members, has both taxa on
the same side of the partition $A|B$.

\begin{cor}
\label{cor:edge_FNJ} If $\delta$ is an $A|B$-consistent dissimilarity
map with respect to a tree $T$, then FNJ applied to $\delta$
will reconstruct a tree $T'$ which contains $A|B$ among its set of
edge-induced splits.
\end{cor}

We conclude with a final comment on $A|B$-consistency and our proof of
Theorem 25.

\begin{figure}[ht]
\label{Fig:chinese}
  \begin{center}
   \includegraphics[scale=0.65]{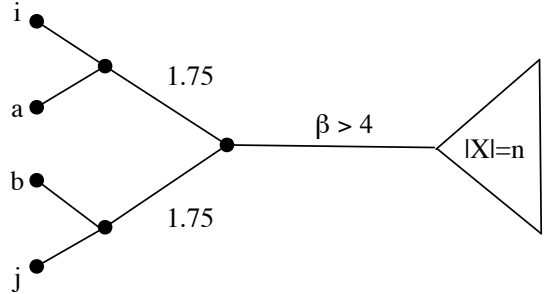}
  \end{center}
\caption{Example for Theorem 34.}
\end{figure}

\begin{thm}\label{thm:chinese_counter}
Let $\delta_T$ be a tree metric and $\delta$ a dissimilarity map whose
$l_\infty$  distance to $\delta_T$ is less than $\frac{\beta}{4}$
where $\beta$ is the length of some edge in $T$. Then it may be
that an intermediate tree produced during the agglomeration steps of the
neighbor-joining algorithm has $l_{\infty}$ distance greater than
$\frac{\beta}{4}$ to any tree metric.
\end{thm}
{\bf Proof}: Consider the phylogenetic tree $T$ in Figure \ref{Fig:chinese}, with
leaf
set $S=X\cup\{i,j,a,b\}$ where $|X|=n$. Suppose that the leaf edges corresponding to
$i,j,a,b$ all have length $\alpha$. The lengths of the other visible
edges, i.e. the ones not belonging to $T|_X$ are as in Figure
\ref{Fig:chinese}. Also suppose that the edges on the
subtree $T|_X$ have total length $\leq\epsilon$, where we will let
$\epsilon$ become arbitrarily small.

Consider the following dissimilarity map $\delta$ with
$||\delta-\delta_T||_{\infty} = 1$:

\begin{enumerate}

\item $\delta(i,j)=\delta_{T}(i,j)-1,$
\item $\delta(p)=\delta_{T}(p)+1$ for $p=(a,b),(a,i),(b,j),$
\item $\delta(p)=\delta_{T}(p)$ for $p=(a,j),(b,i),$
\item $\delta(x,y)=\delta_{T}(x,y)$ for $x,y\in X,$
\item $\delta(i,x)=\delta_{T}(i,x)+1$ for $x\in X,$
\item $\delta(j,x)=\delta_{T}(j,x)+1$ for $x\in X,$
\item $\delta(a,x)=\delta_{T}(a,x)-1$ for $x\in X,$
\item $\delta(b,x)=\delta_{T}(b,x)-1$ for $x\in X.$

\end{enumerate}
Minimizing the  neighbor-joining $Q$ criterion is equivalent to
maximizing
\begin{equation*}
Y_{\delta}(k,l)=2\delta(k,l)+\sum_{t\in S-\{k,l\}} \delta_{k,l}(t)
\end{equation*}
where $\delta_{k,l}(t)=\delta(k,t)+\delta(l,t)-\delta(k,l)$.

First we show that for large enough $n$, small enough $\epsilon$ and with
$\beta>4$, the pair that maximizes $Y$ is $(i,j)$. Note that  for $x,y\in X$, $Y(x,y)=\sum_{t\in \{i,j,k,l\}}
2\delta_{x,y}(t)+O(\epsilon)$, which converges
to a constant as $\epsilon\rightarrow 0$. However, as $n\rightarrow
\infty$, $Y(i,j)\approx 2n\beta$. Therefore for small enough $\epsilon$  and
large enough $n$, $(i,j)$ will dominate any pair $(x,y)\in {X \choose
  2}$. Since $\beta>4$ and $||\delta-\delta_T||_\infty=1$, using a
similar argument as above we can also conclude that the optimum pair must consist of
two leaves from $\{i,j,a,b\}$. 

Finally, we need to show that $Y(i,j)>Y(k,l)$ where either $k$ or $l$ is
equal to $a$ or $b$. Note that for any $k,l \in \{i,j,a,b\}$,
\begin{equation*}
Y(k,l)=2\delta(k,l)+\sum_{t\in \{a,b\}\}}
 \delta_{k,l}(t)+\sum_{t\in X} \delta_{k,l}(t).
\end{equation*}
The first and second summands are constants in $n$, while
the third is composed of $n$ sub-terms which are roughly equal, up to
small variations of size at most $O(\epsilon)$, depending on the
location of $t$ in $T|_X$. Since we can choose $\epsilon$ arbitrarily
small, we can therefore ignore this error.
Now place a fictitious node $v$ at the root of $T|_X$ (the right hand
side of the edge of length $\beta$).Then letting $n\rightarrow \infty$, we see that
asymptotically,
\begin{equation*}
Y(k,l)\approx n \delta_{k,l}(v).
\end{equation*}
Here we extend the definition of $\delta$ to $v$ by extending $\delta_T$
to $v$ in the natural way and defining the error $\delta-\delta_T$ for
pairs involving $v$ in the same way as for other leaves in $X$. 
It is now easy to verify that
\begin{enumerate}
\item
  $\delta_{i,j}(v)=\delta(i,v)+\delta(j,v)-\delta(i,j)=2(\alpha+\beta+1.75)+1+1-(2\alpha+3.5-1)=2\beta+3$,
\item
  $\delta_{i,a}(v)=\delta_{j,b}(v)=\delta(i,v)+\delta(a,v)-\delta(i,a)=2(\alpha+\beta+1.75)+1-1-(2\alpha+1)=2\beta+2.5$,
\item
  $\delta_{j,a}(v)=\delta_{i,b}(v)=\delta(j,v)+\delta(a,v)-\delta(j,a)=2(\alpha+\beta+1.75)+1-1-(2\alpha+3.5)=2\beta$,
\item
  $\delta_{a,b}(v)=\delta(a,v)+\delta(b,v)-\delta(a,b)=2(\alpha+\beta+1.75)-1-1-(2\alpha+3.5+1)=2\beta-3$.
\end{enumerate}

Therefore asymptotically, $(i,j)$ will be collapsed in the first step of neighbor-joining applied to $\delta$. 

Now consider the reduced distance matrix
$\delta':S'\times S'\rightarrow \mathbb{R}$, where the leaves $i$ and
$j$ are replaced by a new leaf $u$. That is, $S'=S-\{i,j\}\cup\{u\}$. We restrict our attention to the set
$\{a,b,u,x\}$ for an arbitrary leaf of $x$ of $X$. Since the total
length of $T|_X$ is $O(\epsilon)$, we can in fact approximate expressions
involving $\delta(\dot,x)$ by $\delta(\dot,v)$ up to $O(\epsilon)$
error. We do so for the sake of simplicity. 

Simple calculations now give
$$\delta'(a,u)=\delta'(b,u)=2\alpha+2.25,$$
$$\delta'(a,x)=\delta'(b,x)=\alpha+\beta+.75,$$
$$\delta'(a,b)=\delta(a,b)=2\alpha+3.5,$$
$$\delta'(u,x)=\alpha+\beta+2.75,$$
and thus:
$$\delta'(a,u)+\delta'(x,b)=\delta'(b,u)+\delta'(x,a)=\delta'(x,u)+\delta'(a,b)-4.25=\beta+3\alpha+3.$$

Now suppose that there is some additive tree metric $\mu$ such that
$||\delta'-\mu||_\infty\leq 1$. Then by adding this to the above
equality we obtain that
$$\mu(x,u)+\mu(a,b)-\mu(b,u)-\mu(x,a)\geq 4.25-4>0$$ and similarly
$$\mu(x,u)+\mu(a,b)-\mu(a,u)-\mu(x,b)\geq 4.25-4>0.$$ This contradicts
the four point condition necessary for $\mu$ to be a tree metric. 
Therefore, we have given an example of a dissimilarity map $\delta$ and a
tree metric $\delta_T$ with an edge of length $\beta$, such that $||\delta -
\delta_T||_{\infty}<\frac{\beta}{4}$, and yet the $l_\infty$ distance of the
reduced dissimilarity map after the first agglomeration step from \emph{any} tree
metric is greater than $\frac{\beta}{4}$. \qed

The significance of Theorem 34 is that it shows that an inductive proof
of Atteson's conjecture is not possible without relaxing the
hypothesis. Thus, the partial result of \cite{Xu2005} and the proof
of \cite{Dai2006} are incorrect. 
At the same time, Theorem \ref{thm:chinese_counter} also identifies an
undesirable property of neighbor-joining which is very common among
greedy optimization algorithms. We hope that further
investigations in this direction can yield more robust versions of the algorithm.

\section{Simulation results and conclusion}

\begin{figure}[]\label{fig:graph}
\begin{center}
\includegraphics[scale=0.5]{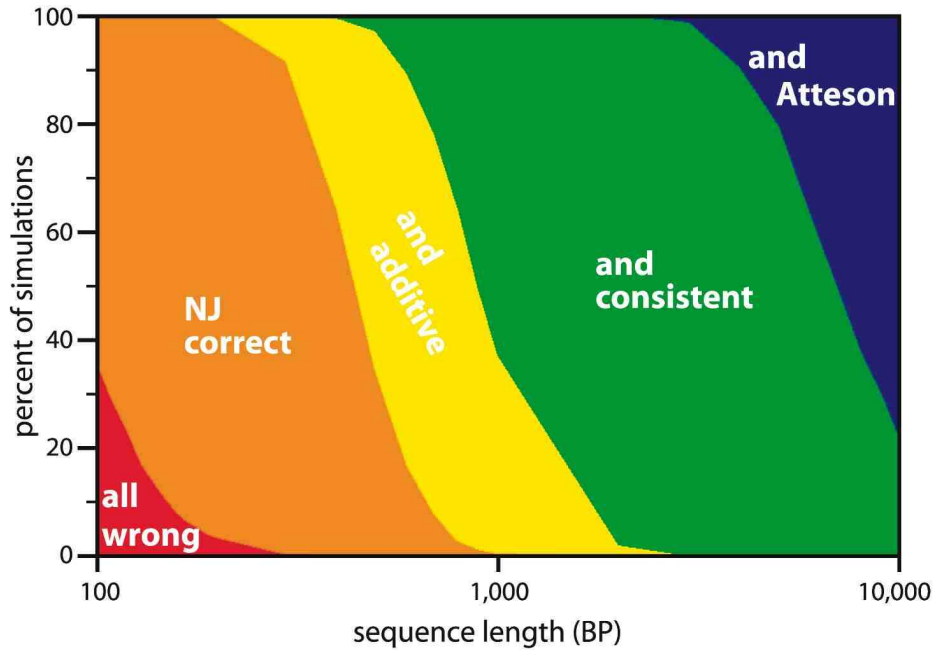}
\caption{Conditions satisfied as a function of sequence length.  35
  trees with 20 taxa each were simulated 100 times for 28 different
  sequence lengths. 100 alignments were generated for each tree.
  The figure shows, for each of the 9800000 dissimilarity maps generated
  from the simulations, whether neighbor-joining reconstructed the
  correct tree, and whether additivity, consistency and Atteson's
  criterion were satisfied. Note that the x-axis is logarithmic.}

\end{center}
\end{figure}

We performed a series of simulations to test how frequently Theorem 16
explains the success of the neighbor-joining algorithm.
Trees with 20 taxa were generated by agglomerating pairs at
random. We generated 35 such trees and set their edge lengths
to 0.1.  We then used \texttt{seq-gen} \cite{seq_gen} to build 100
alignments with the Jukes Cantor model for each of 28 sequence
lengths between 100 and 10,000 base pairs.  From these we obtained
dissimilarity maps  using the \texttt{dnadist} program
from the PHYLIP package \cite{phylip}.  Our Java API then computed the
$w$-matrix for each dissimilarity map, tested the consistency and
additivity conditions against the true tree, checked to see if the
dissimilarity map satisfied Atteson's criterion (Theorem 2), and
computed the
neighbor-joining tree.  The results are summarized in Figure
\ref{fig:graph}.

We note that our conditions of additivity and
consistency are satisfied for sequence lengths an order of magnitude
smaller than required for Atteson's criterion to hold.
Moreover, even when the additivity and consistency
conditions are not satisfied for every quartet,
they do hold true for upwards of 99\% and 94\% of quartets respectively,
even at sequence length 100 (Figure 7).  Hence, applying an averaging
argument
similar to the one we employed in the proof of Atteson edge radius
conjecture, we
may obtain ``on average'' conditions that explain even more of the cases
where neighbor-joining succeeds.

\begin{figure}[]\label{fig:graph2}
  \begin{center}
\includegraphics[scale=0.5]{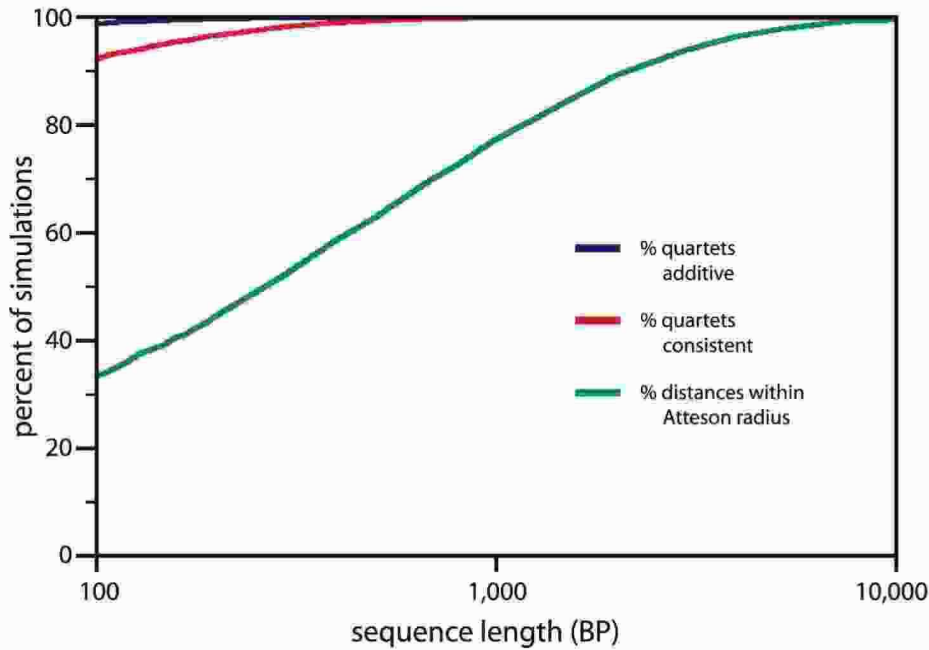}
\caption{Details on the percent of quartets satisfying the additivity
  and
consistency conditions when neighbor-joining succeeds. The green curve
shows the number of distances within the Atteson radius.}
\end{center}
\end{figure}

\section{Acknowledgments}
Radu Mihaescu was supported by a National Science Foundation graduate
fellowship, and partially by the Fannie and John Hertz
Foundation. Lior Pachter was partially supported by
NIH grant R01HG2362 and NSF grant CCF0347992. Dan Levy was supported by
NIH grant GM68423.

\providecommand{\bysame}{\leavevmode\hbox to3em{\hrulefill}\thinspace}
\providecommand{\MR}{\relax\ifhmode\unskip\space\fi MR }
\providecommand{\MRhref}[2]{%
  \href{http://www.ams.org/mathscinet-getitem?mr=#1}{#2}
}
\providecommand{\href}[2]{#2}

\end{document}